\documentclass{elsart}
\usepackage{graphicx}
\newcommand{\req}[1]{(\ref{#1})}
\newcommand{\be}{\begin{equation}}
\newcommand{\ee}{\end{equation}}
\newcommand{\bea}{\begin{eqnarray}}
\newcommand{\eea}{\end{eqnarray}}

\newcommand{\cro}[1]{\left[#1\right]}

\newcommand{\avg}[1]{\langle{#1}\rangle}

\newcommand{\BE}{\begin{eqnarray}}
\newcommand{\EE}{\end{eqnarray}}
\newcommand{\BEn}{\begin{eqnarray*}}
\newcommand{\EEn}{\end{eqnarray*}}
\newcommand{\barr}{\begin{array}}
\newcommand{\earr}{\end{array}}

\newcommand{\bit}{\begin{itemize}}      
\newcommand{\eit}{\end{itemize}}
\newcommand{\bc}{\begin{center}}
\newcommand{\ec}{\end{center}}
\newcommand{\ben}{\begin{enumerate}}    
\newcommand{\een}{\end{enumerate}}

\begin{document}

\begin{frontmatter}
\title{Limit order market analysis and modelling:\\on an universal cause for over-diffusive prices}
\author{Damien Challet and Robin Stinchcombe}
\address{Theoretical Physics, 1 Keble Road, Oxford OX1 3NP, United
Kingdom\\{\tt challet@thphys.ox.ac.uk~~~r.stinchcombe1@physics.ox.ac.uk}}
\date{\today}


\begin{abstract}
We briefly review data analysis of the Island order book, part of
NASDAQ, which suggests a framework to which all limit order market models
should comply. Using a simple exclusion particle model, we argue that
short-time price
over-diffusion in limit order markets is due to the non-equilibrium
of order placement, cancellation and execution rates, which is an
inherent feature of real limit order markets.
\end{abstract}
\end{frontmatter}


By contrast with usual data on volume and transaction price of a
set of stocks or foreign exchanges~\cite{BouchaudPotters,MantegnaStanley,Daco}, limit order markets provide
instantaneous information about unfilled orders. In the language of Physics, one has 
access to an additional dimension, the price. The analysis and the modeling of
such markets is therefore more complex.
Economics literature has focused on the design of limit order
markets (see~\cite{Coppejans,Domowitz}), asking for instance why and when limit orders may be preferred
by traders~\cite{Cohen} and introducing sophisticated models based on
equilibrium prices~(see for instance \cite{Soph}), while  physicists
recently begun to analyze data and to play with
dynamics stochastic toy-models where the price evolution is a
consequence of stochastic order arrival~\cite{Bak,Kogan,Maslov,Maslov2,CS01,Farmer,BouchaudLimit,Gunter,CS02,Farmer2}.

Some definitions first. A limit order is characterized by
three properties: its price $p$, its size $m$ and its lifetime
$\tau$. When it is placed into the order book of a given stock, it
publishes the wish to buy (or sell) $m$ units of this stock at the
predefined price $p$. If there is already a sell order (buy order) in the
book at a lower or equal price, 
both orders will be automatically executed, partly or fully,
depending on the size of matching opposite order(s). If this is not
the case the new order will wait in the book until a compatible
opposite order is placed or until its lifetime is over. This implies that at
any time, there is a spatial distribution of unfilled buy and sell orders. The
best prices are defined as the largest buy price and the lowest sell
price. The difference between the two is called the bid-ask spread.
As often emphasized, submitting a limit order is a trade-off
between the advantage of obtaining a fixed price, and the disadvantage
of not knowing precisely how long the order will have to
wait. 

We collected data from Island.com, a subpart of the NASDAQ. Each order
has a unique ID number, which allowed us to keep track of their
individual fates. Only the 15 best orders of each type were public at
any given time, but we managed to find information about 80\% of the
orders. The major problem with this kind of data is the uncertainty in
time it causes for orders that are not often listed in the 15 best ones.

\begin{figure}
\centerline{\includegraphics[width=8cm]{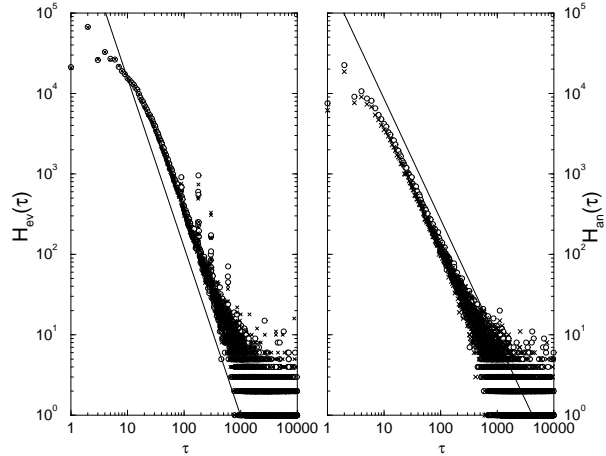}}
\caption{Lifetime histogram for bid (x) and ask orders (circles) of all orders of all stocks and all days. The left graph plots  $H_{\rm ev}$, the histogram of cancelled orders (straight-line: power-law with a $-2.1$ exponent). The right graph plots $H_{\rm an}$, the histogram of annihilated orders (straight-line: power-law with a $-1.5$ exponent).} 
\label{lifetime}
\end{figure}

We measured first order lifetime distribution, shown in
Fig.~\ref{lifetime}. For both the cancelled and the executed orders,
we find that its tail  is well fitted by a power-law $P(\tau)\sim
\tau^{-\alpha}$ with $\alpha\simeq2.1$ for cancelled orders and $\alpha\simeq1.5$
for annihilated orders for $\tau\le1000$ and then tend to broaden. This
last effect is clearly due to the nature of our data: it attributes a too long
lifetime to orders that are seldom seen, i.e. far from best prices. 
 Characteristic lifetimes of $90$, $120$ and
$180$ seconds clearly appear on left panel, while the bulk of the
distribution is due to active order canceling, that is, to traders watching the evolution of the market and cancelling their orders before their predefined lifetime is reached. It is tempting to
relate this power-law to the dependence of the
evaporation rate on the relative position, which is also a
power-law~\cite{Potters2}. 

\begin{figure}
\centerline{\includegraphics[width=8cm]{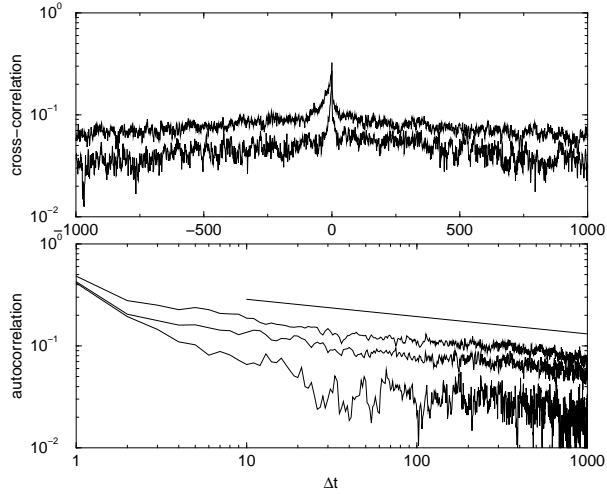}}
\caption{Bottom graph: autocorrelation of the bid rates $\delta$ (top), $\eta$ (middle) and $\alpha$ (bottom) measured with $\Delta t=1$ s the dashed line has a -0.17 exponent. Upper graph: crosscorrelations $\avg{\delta(t)\eta(t+\Delta t)}$ (top) and $\eta$, $\avg{\delta(t)\alpha(t+\Delta t)}$} 
\label{corrs}
\end{figure}

We next turn to the measure of deposition (order placement),
anihilation (market orders), and
evaporation (order cancellation) rates, denoted $\delta$, $\alpha$ and
$\eta$ respectively, which highly fluctuate during a trading day. 
Note that deposition and evaporation are
proportional as a first approximation~\cite{CS01}. All these rates have algebraically
decreasing cross-correlation, which is expected for annihilation rates, as it corresponds to that of the volume of transaction~\cite{MantegnaStanley}. In addition, the asymmetry in the
deposition-evaporation cross-correlation function implies that
evaporation triggers deposition: a trader that cancels an order is
likely to wish to place it again~\cite{CS01}. However, we could
not see trace of massive order diffusion, which is the fundamental assumption
of the family of models found in Refs~\cite{Bak,Kogan}.

We found a linear dependence of the relative deposition width on the
bid-ask spread, which can be explained by orders deposited at half of
the spread, or directly at the tick adjacent to the best opposite best price.

The above observations leave us with the following framework~\cite{CS01} that should
be the basis of any stochastic particle model of limit order markets:
\ben
\item  bulk-deposition relative to the best price of each type of orders, with 
some spatial distribution. According to Refs~\cite{BouchaudLimit,Farmer},
it is a power law;
\item spread-deposition with some spatial distribution;
\item market orders, or cross-deposition; they can be included into the previous ingredient.
\item cancellation of orders.
\een

The last ingredient is crucial in these kind of model, and was introduced
in~\cite{CS01}. Its presence ensures for instance that the number of orders does
not diverge with time, and that the price evolution is diffusive for
large times. Indeed, all particle
models that assume fixed rates display under-diffusive
prices~\cite{Bak,Maslov,CS01,BouchaudLimit,Farmer}, that is,
$\sum_{t=1}^T\delta p_t\sim T^{\beta}$ with $\beta<1/2$, where $\delta p_t$ is the price increment at time $t$, 
with a crossover to
diffusive prices ($\beta=1/2$) as soon as the orders can
evaporate~\cite{CS01}. Note also that evaporation is responsible for
the crossover from $\beta=2/3$ and $\beta=1/2$ in the model proposed
by  Ref.~\cite{Gunter}.
However, the rates depend on time in real markets in such a way that
the rate imbalance, that can be defined as the difference of
$\delta-\alpha-\eta$ between asks and bids, is never zero. In particle models,
this induces trends which lead to
overdiffusive behaviours ($\beta>1/2$)~\cite{CS02}. The possibly simplest model with time
varying rates is an particle exclusion model where the rates are
changed all at the same time~\cite{Gunter,CS02}. In exclusion models of particles, only
 only one particle can stand on a given site at a given time. Some of
these models are exactly solvable~\cite{Derrida,Stinchcombe}. They provide an ideal playground to test ideas
on models of limit order markets.

The simplicity of these kind of models makes it possible to compute exactly some
quantities, such as the Hurst exponent $\beta=2/3$ in~\cite{Gunter}. An
analogy between price changes in the model of Ref~\cite{CS02} and a generalized Ising
spin $1-d$ model qualitatively explains the observed overdiffusive
behavior.

\begin{figure}
\centerline{\includegraphics[angle=270,width=8cm]{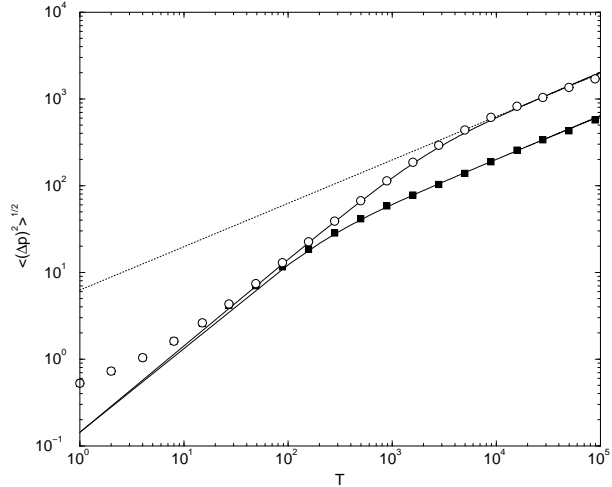}}
\caption{Hurst plot of the price increments for probabilities of rate change $p=0.001$ (circles) and $0.01$
(squares) . Deposition rates were drawn between 0 and 1, and
annihilation rates between $0$ and $0.2$. No evaporation. $10^6$ iterations. The dotted line represents normal random walk ($\beta=1/2$) and continuous lines are obtained from Eq.~\req{H}.} 
\label{over}
\end{figure}

Here, we repeat the argument of Ref~\cite{CS02} by showing
numerically that time-varying rates are the key to understanding 
overdiffusive price behaviour in financial markets. We
consider totally random and exponentially correlated rates. The only assumption is
that all rates are redrawn from an uniform distribution $[0,R]$ with
fixed probability $p$ at each time
step, with $R=1$ for deposition and $R<1$ for
annihilation.\footnote{Considering the independent changes of each rate
with some probability at each time step leads to the same results.}
 As shown in~\cite{CS02}, a set of rates defines a price drift, hence
a trend. Although the rates
are shortly correlated, this is enough to produce overdiffusive prices, as shown by
Fig.~\ref{over}, and is explained by the fact that the rates stay
constant over a period of time. This induces some memory in the rates themselves, consistent
with the algebraical decay of rates autocorrelation (although here the
decay is much faster). Neglecting bid-ask spread bounces and retaining only ballistic moves due to the drift, it is straightforward  to find that price fluctuations are given by 
\be\label{H}
(\sum_{t=1}^T\delta p_t)^2\propto T+\frac{2(1-p)}{p^2}\,\cro{pT-1+(1-p)^T},
\ee
which is correct for $T$ large enough (see fig. \ref{over}). The diffusive behaviour for small $T$ can be attributed mostly to bid-ask spread bounces and other sources of noise that where neglected in our derivation.
From a practical point of view, we see no reason why the bid and ask rates should be equal at any
time in real markets. This would require for instance that traders could react
simultaneously to the state of the order book or to an information.
Heterogeneous reaction times can be considered as a simple cause for the imbalance. 
Therefore, we argue that the short-term overdiffusive behaviour of prices in
limit order markets is quite possibly due to the unavoidable temporal imbalance of
these rates. This of course does not explain long term (several
months) overdiffusive prices~\cite{Daco,MantegnaStanley}. A better
explanation may be herding behaviour~\cite{ContBouchaud,Lux}, or
particular economic situations.

In conclusion, we believe that Physics can contribute to the
understanding of limit order markets, by asking different questions
than Economics. For us, these markets are similar to non-equilibrium
particle lattice systems, where the variability of
deposition/annihilation bid/ask rates is the cause of short term overdiffusive
price evolution, and must be incorporated into current existing
models, as in~\cite{CS02}.

We would like to thank Th. Bochud, J.-Ph Bouchaud, J. Iori, M. Potters, R. Rajesh for stimulating discussions.



\end{document}